\documentclass{aa}
\usepackage{graphics}
\begin{document}

   \thesaurus{11     % A&A Section 11: Galaxies
              (11.19.3;  % Galaxies: starburst,
	       11.01.1;  % Galaxies: abundances,
               11.05.2;  % Galaxies: evolution,
               11.06.1)} % Galaxies: formation.

\title{The abundance of nitrogen in starburst nucleus galaxies
\thanks{Based on observations obtained at the 193cm telescope of
Observatoire de Haute-Provence, operated by INSU (CNRS)}}

\author{R. Coziol\inst{1}\inst{,2}
\and
R. E. Carlos Reyes\inst{3}\inst{,4}
\and
S. Consid\`ere\inst{5}
\and 
E. Davoust\inst{6}
\and
T. Contini\inst{7}
}

\institute{Observat\'orio Nacional, Rua Gal.  Jos\'e Cristino,
77 -- 20921-400, Rio de Janeiro, RJ., Brasil
%%email: coziol@maxwell.on.br
\and
PRONEX/FINEP -- P. 246 -- 41.96.0908.00
\and
Laborat\'orio Nacional de Astrof\'{\i}sica - LNA/CNPq,
CP 21, 37500--000 Itajub\'a, MG, Brasil
%%email: rafael@lna.br
\and
Seminario Permanente de Astronom\'{\i}a y Ciencias Espaciales, FCF, UNMSM, L\'{\i}ma, Per\'u
\and
UPRES-A CNRS 6091, Observatoire de Besan\c{c}on, B. P. 1615, F-25010 Besan\c{c}on cedex, France
%%email: sc@obs-besancon.fr
\and
UMR 5572, Observatoire Midi--Pyr\'en\'ees, 14 avenue E. Belin, F-31400 Toulouse, France
%%email: davoust@obs-mip.fr 
\and
European Southern Observatory, Karl- Schwarzschild-Str. 2, 85748, Garching bei Munchen, Germany
%%email: tcontini@eso.org
}
%%\offprints{R. Coziol, coziol@maxwell.on.br}
\offprints{E. Davoust, davoust@obs-mip.fr}

   \date{Received ; accepted }

   \maketitle

\begin{abstract}

We show that the excess of nitrogen emission observed in a large sample of 
Starburst Nucleus Galaxies (SBNGs) can only be explained
at a given metallicity by an overabundance of nitrogen with respect to normal 
H\,{\sc ii} regions in 
the disks of late--type spirals.  The N/O ratios in the SBNGs are comparable 
to the values found in the bulges of normal early--type spirals, which 
suggests that what we observe could be the main production of nitrogen in the 
bulges of these galaxies. 

The variation of the N/O ratio as a function of 
metallicity in SBNGs follows a $primary + secondary$ 
relation, but the increase of nitrogen does not appear as 
a continuous process. In SBNGs, nitrogen is probably produced by different 
populations of intermediate-mass stars, which were formed during past 
sequences of bursts of star formation. This assumption pushes the origin of 
the main bursts 2--3 Gyrs back in the past. On a cosmological scale, this time 
interval corresponds to redshifts $z \sim 0.2-0.3$, where a significant 
increase of star formation activity occurred. The origin of the SBNG 
phenomenon would thus have cosmological implications, it would be related to a 
more active phase of star formation in the Universe sometime in its recent 
past. 
   
\end{abstract} 


\section{Introduction} 

Starburst galaxies can be grouped in two main families: the H\,{\sc ii} 
galaxies and the Starburst Nucleus Galaxies (SBNGs). The former are small-mass 
and metal-poor galaxies, while the latter are more massive and metal-rich (for 
a complete definition of starburst galaxies, see Coziol et al. 1998a). The 
astronomical community recognized quite early the peculiar nature of H\,{\sc 
ii} galaxies and many studies have been undertaken to determine their 
characteristics. The origin of SBNGs, on the other hand, was thought to be 
more straightforward to understand. As a consequence, the basic 
characteristics of these galaxies are not as well determined and their nature 
is still in debate.  

For example, it was believed some twenty years ago that SBNGs were 
``old'' galaxies which were rejuvenated by inflow of matter resulting from 
interactions with other galaxies (Huchra 1977; Tinsley \& Larson 1978). But 
subsequent studies have shown that starbursts generally do not reside in high 
galaxy density regions (Salzer et al. 1989; Hashimoto et al. 1998; Coziol et 
al.\ 1998b, 1999), where conditions for interaction should be more favorable, 
and that only a fraction (between 25\% and 30\%) of the massive SBNGs have 
obvious luminous companions (Keel \& van Soest 1992; Gallimore \& Keel 1993; 
Coziol et al. 1995; Coziol et al. 1997a, 1999; Contini et al. 1998). 

The frequency of interacting galaxies among SBNGs may be higher however, 
if we assume that the morphological peculiarities often observed in 
these galaxies, are weak traces of past interactions (Keel \& van Soest 1992; 
Coziol et al. 1995, Barth et al. 1995) and that, consequently, many SBNGs are the
remnants of merging galaxies (Coziol et al. 1998c). 
The assumption that SBNGs result from accretions of smaller galaxies is 
supported by several facts : they are 
less chemically evolved than normal galaxies with similar morphologies and 
comparable luminosities (Coziol et al. 1997b),
they are predominantly early--type spiral galaxies 
(Coziol et al. 1997a; 1998a) and follow a luminosity--metallicity 
relation similar to that of elliptical galaxies (Coziol et al. 1998c).

The merger hypothesis, on the other hand, does not 
completely explain the starburst phenomenon and it is probably 
also necessary to assume internal regulating star formation mechanisms, in
order to 
extend the total lifetime of the starburst over a longer period than is 
usually assumed based on the dynamical time scale of a gravitational 
interaction between galaxies (Searle \& Sargent 1972; Gerola et al. 1980; 
Kr\"{u}gel \& Tutukov 1993).  Such mechanisms would explain 
why SBNGs experienced more than one burst over the last few Gyrs 
(Gonz\'alez~Delgado et al. 1995; Coziol 1996).   
They would also explain why these galaxies do not produce supernovae at 
higher rates than normal ones (Turatto et al. 1989; Richmond et al.
1998; Gonz\'alez~Delgado et al. 1999) 

In principle, one could gain a better understanding of the nature and origin 
of SBNGs by drawing a more complete picture of their chemical evolution. But 
this is a difficult task, which is further complicated by our relative 
ignorance of how the chemical evolution of normal spiral galaxies proceeds. 
The main difficulty is observational: building a reasonable scenario of the 
chemical evolution of galaxies generally requires observations of H\,{\sc ii} 
regions over a large range of wavelengths. For local galaxies, for example, UV 
observations are required for determining the abundance of carbon (Kobulnicky 
\& Skillman 1998), while observations in the near infrared are needed for 
estimating the abundance of sulphur (Garnett 1989). In metal-rich spiral 
galaxies, an additional difficulty resides in the non observability of 
important nebular lines, like [O\,{\sc iii}] at 4363 \AA, without which the 
electron temperature of the gas cannot be determined directly. 

Many efforts have been devoted over the years to develop functional empirical 
techniques to circumvent the problems related to the determination of chemical 
abundances in galaxies (Pagel et al. 1979; Edmunds \& Pagel 1984; Edmunds 
1989; Skillman 1989). Recently, Thurston et al. (1996; hereafter TEH) have 
extended these techniques and devised a new method for estimating nitrogen 
abundances in metal-rich galaxies.  In this paper, we take advantage of this 
recent advance to determine the abundance of nitrogen in SBNGs. 
  
The organization of our paper is as follows. In Section~2, we identify the main
source of excitation in our sample of SBNGs, based on their location in
three standard diagnostic diagrams. We then discuss, in Section~3, the legitimacy of 
applying to SBNGs the empirical method designed for ``normal'' H\,{\sc ii} 
regions to determine their chemical abundances. In the SBNGs, we took great 
care in verifying that no other sources of excitation like a hidden AGN 
is present, which could complicate or even impede the abundance 
determination in these galaxies. Having verified these points, we proceed 
in Section~4 to estimate the oxygen and nitrogen abundances in a large sample 
of SBNGs, comparing these abundances with
those of normal H\,{\sc ii} regions in spiral galaxies. In Section~5, we 
discuss two different scenarios for producing nitrogen in SBNGs and 
examine their consequences for the origin of the bursts in these galaxies. A 
summary of our most salient results is presented in Section~6.   

\section{The location of SBNGs in standard diagnostic diagrams} 

The optical spectrum of a starburst galaxy is usually quite distinct from that 
of an AGN\footnote{
We use the word AGN as it means, namely Active Galactic Nucleus. We
do not presuppose anything as to the origin of this activity, only that it
involves something more than what is usually observed in
normal H\,{\sc ii} regions: either a black hole or shocks.}. 
It resembles that of an H\,{\sc ii} region in our Galaxy, where, as 
we know, the gas is photoionized by young O and B stars (Searle \& Sargent 
1972; Sargent 1972; Huchra 1977; French 1980). This conclusion applies to all 
the types of starburst galaxies known today (Salzer et al. 1989; Kim et al. 
1998; Coziol 1996; Coziol et al. 1998a), including those where the burst is 
located mostly in the central regions (Weedman et al. 1981; Balzano 1983). 

But, because the physical conditions in the nuclei of galaxies are still 
poorly understood, one may wonder what kind of pathological phenomena the 
evolution of a powerful starburst in the center of a galaxy could develop 
(Weedman 1983; Terlevich et al. 1991).  This concern may partly be justified 
by observations. For example, the presence of low-intensity non-stellar 
activity in the center of normal galaxies is frequently suggested (Heckman 
1980; Peimbert \& Torres-Peimbert 1981; V\'eron et al. 1981; Rose \& Cecil 
1983; Filippenko \& Sargent 1985) and, in AGNs, traces of star formation 
occurring simultaneously with an active nucleus are sometimes noted (Shields 
\& Filippenko 1990; Lonsdale et al. 1993; V\'eron et al. 1997; 
Gonz\'alez~Delgado et al. 1997).  However, none of these observations constitutes 
definite evidence that an evolutionary connection between starburst and AGN 
exists, and, even less so, explains how this evolution may proceed.  

%----------------------------------------------------------- S_vib
\begin{figure*}
\hskip -2cm 
\vskip 1cm
\resizebox{15cm}{!}{\includegraphics{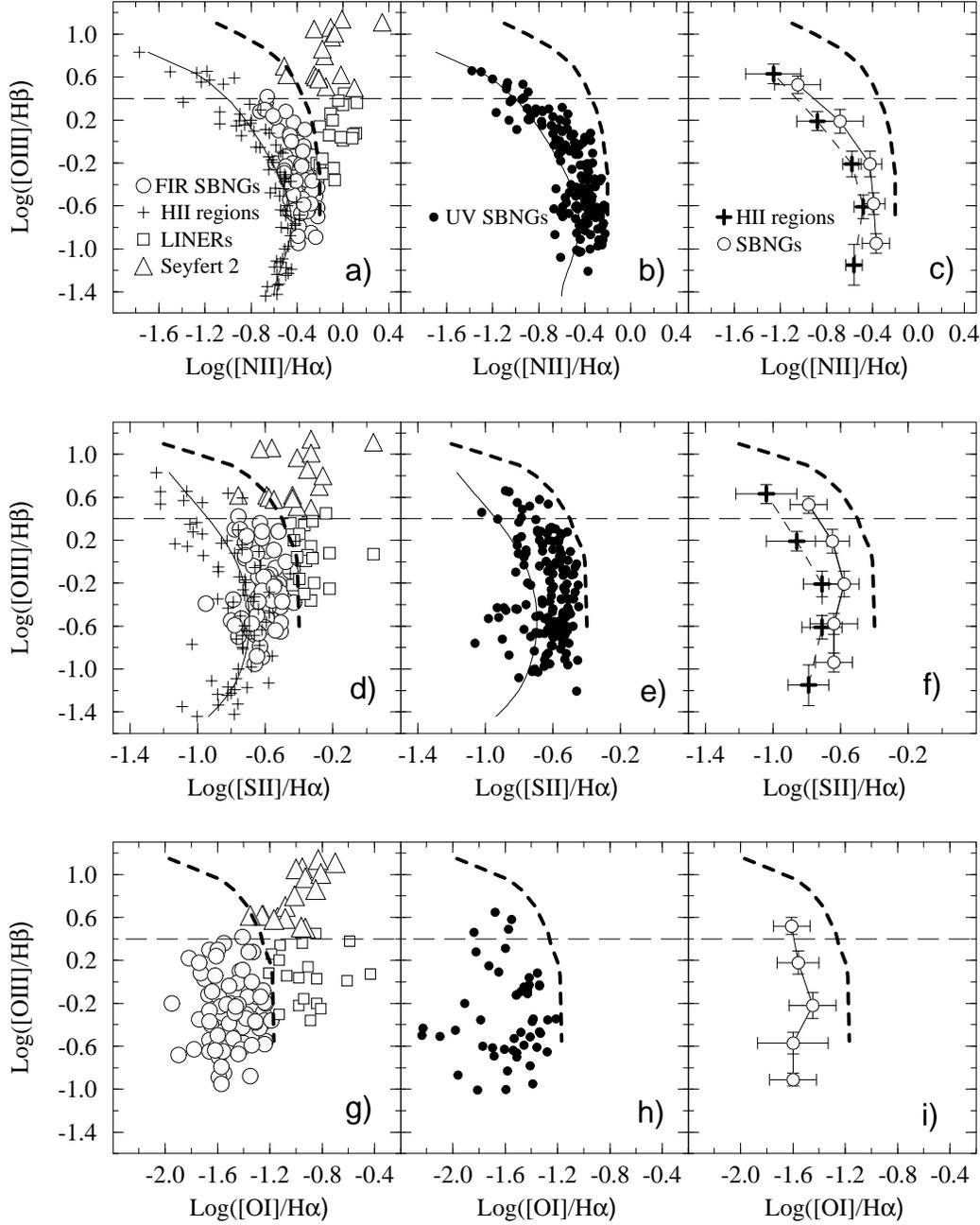}}
\hfill
\centering
\vskip -2.cm
\parbox[b]{18cm}{
\caption[h]{Three standard diagnostic diagrams of line ratios used to 
determine the dominant source of excitation in emission-line galaxies (BPT, 
VO). In these diagrams, the thick dashed curves are the empirical separations 
between starbursts and AGNs (Seyfert 2s and LINERs) proposed by VO. The thin 
horizontal dashed lines separate low--excitation from high--excitation 
galaxies (Coziol 1996). The thin continuous curves are regressions on the MRS 
sample of normal H\,{\sc ii} regions.  The other symbols are explained in a), 
b) and c).  AGNs are clearly separated from normal H\,{\sc ii} 
regions and SBNGs in a), d) and g). The SBNGs generally have lower levels of 
excitation than Seyfert 2s [(O\,{\sc iii]}/H$\beta < 0.4$) and lower [N\,{\sc 
ii}]/H$\alpha$, [S\,{\sc ii}]/H$\alpha$ and [O\,{\sc i}]/H$\alpha$ ratios than 
LINERs. In c), f) and i), we compare the mean values of [N\,{\sc ii}], 
[S\,{\sc ii}] and [O\,{\sc i}] estimated in excitation-level bins of 0.4 dex. 
In general, the SBNGs trace a similar sequence to normal H\,{\sc ii} regions, 
except for a slight excess of emission.  As the level of excitation increases, 
the emission of nitrogen decreases while those of sulphur and oxygen stay 
roughly constant} 
\label{Fig.1}} 
%\label{PBT-VO}}
\end{figure*}
%
%______________________________________________________________

Obviously, if a source of excitation other than star formation is active in 
SBNGs, this will greatly complicate abundance determinations. It is thus 
essential to first establish the dominant excitation source (O and B stars 
or an AGN) present in these galaxies. This is usually done empirically: 
one constructs diagnostic diagrams using several ratios of emission lines 
present in the optical spectra (Baldwin et al. 1981; Veilleux \& Osterbrock 
1987, hereafter BPT and VO). Over the years, the use of diagnostic diagrams 
has been recognized as the most efficient method for classifying the different 
types of activity encountered in emission-line galaxies (Veilleux et al. 1995; 
V\'eron et al. 1997; Gon\c{c}alves et al. 1999). This is the method that we 
applied to identify the main source of excitation in our different samples of 
galaxies. 

Our sample of UV-bright 
SBNGs was composed originally of 208 H\,{\sc ii} regions 
observed along the bars of 75 Markarian barred galaxies (see Contini 1996, 
Contini et al. 1998 and Consid\`ere et al. 1999 for details). From this 
sample, we rejected 48 H\,{\sc ii} regions because of their ambiguous 
classification in the three diagnostic diagrams. This same criterion was used 
by Veilleux et al. (1995) to build their sample of FIR-bright SBNGs. This 
criterion allows one to eliminate the so called ``transition'' galaxies, whose 
nature is not well determined (V\'eron et al. 1997; Gon\c{c}alves et al. 1999; 
Hill et al. 1999; Doyon et al. 1999) \footnote{These ``transition'' types are 
probably not a new kind of emission--line galaxies, but cases of misclassified 
Seyfert or LINER galaxies with circumnuclear star formation (see Gon\c{c}alves 
et al. 1999).}. 

For comparison, we used the sample of 83 FIR-bright SBNGs of Veilleux et al. 
(1995). We also used the 70 H\,{\sc ii} regions observed by McCall et al.
(1985; hereafter MRS) to represent normal H\,{\sc ii} regions. In all 
the samples, the line ratios were corrected for internal reddening using 
Balmer decrements. 

The three classical diagrams of [O\,{\sc iii}]$\lambda5007$/H$\beta$ (the 
excitation level) as a function of [N\,{\sc ii}]$\lambda6584$/H$\alpha$, 
[S\,{\sc ii}]$\lambda(6717+6731)$/H$\alpha$ and [O\,{\sc 
i}]$\lambda6300$/H$\alpha$ for the various samples are shown in Fig. 
\ref{Fig.1}. Note that, for the sake of clarity, these line ratios will 
hereafter be written without the wavelengths. AGNs (Seyfert 2s and LINERs) are 
clearly separated from H\,{\sc ii} regions and SBNGs (Fig. \ref{Fig.1}a, d and 
g). In these diagrams, the H\,{\sc ii} regions trace a sequence where the 
decreasing excitation is due to the increasing oxygen abundance (MRS; Dopita 
\& Evans 1986). In general, the SBNGs trace a similar sequence, but slightly 
displaced towards higher emission-line ratios (Coziol et al. 1997a). 

Despite being slightly more luminous in the infrared, the FIR-bright SBNGs show 
optical spectral characteristics similar to those of the UV-bright SBNGs. This 
is consistent with the similarities observed between these two types
of SBNGs in the FIR and suggests that the 
two selection criteria (UV- or FIR-bright) define mostly the same kind of 
galaxies (Coziol et al. 1998a).  A Kolmogorov--Smirnoff test performed on 
[N\,{\sc ii}] suggests, with a confidence limit greater than 99\%, that the 
FIR- and UV--bright SBNGs come from the same population. The same test 
yields similar conclusions for [O\,{\sc i}] and [S\,{\sc ii}], but only with 
confidence limits greater than 90\% and 80\% respectively. 
 
The [O\,{\sc i}] line is rarely observed in normal H\,{\sc ii} regions (BPT), 
and it was not measured in the MRS sample. When it is measurable in starburst, its 
intensity is always weaker than in AGNs. This is because the 
[O\,{\sc i}] line is produced only in regions of partial ionization which are 
much more extended in AGNs (VO). In our sample of UV--bright SBNGs, only a 
small fraction ($\sim$ 30\%) of the galaxies presents this line.  We compare 
the few [O\,{\sc i}]/H$\alpha$ ratios observed in the SBNGs of our sample in 
Fig. \ref{Fig.1}h and i. Log([O\,{\sc i}]/H$\alpha$) is on average $\sim\ -1.6$,  
comparable to what is observed in the sample of FIR-bright SBNGs of Veilleux 
et al. (1995) and consistent with values found in normal H\,{\sc ii} regions 
(BPT, VO). 

We conclude, on the basis of the three standard diagnostic diagrams, that the 
main source of excitation of the gas in SBNGs is O and B stars, like in normal H\,{\sc ii} regions
(BPT; VO; Gon\c{c}alvez et al. 1998; Kim et al. 1998).

As can be seen in Fig.~\ref{Fig.1}, even if the gas in SBNGs is ionized by
O and B stars, the spectra of these galaxies do show, however, an excess of nitrogen and sulphur 
emissions with respect to H\,{\sc ii} regions in the disk of normal spirals 
(the MRS sample).  Before proceeding with the determination of abundances in SBNGs, 
we will now discuss the possible causes of this phenomenon. 
Note that because we do not have enough information to estimate 
the abundance of sulphur in our samples, our discussion will focus on nitrogen.  

\section{The origin of the excess of nitrogen emission in SBNGs} 

An excess of nitrogen emission in the spectra of galaxies (usually normal ones) has 
been observed before (see Stauffer 1982, and references therein) and various 
hypotheses have been proposed to explain it (e.g. Kennicutt et al. 1989).  The 
simplest hypothesis, which can apply to SBNGs, is that these galaxies 
possess an overabundance of nitrogen with respect to normal disk H\,{\sc ii} 
regions (Stauffer 1982).  Such an overabundance 
has already been observed in the prototypical SBNG NGC 7714 (Gonz\'alez~Delgado et al. 1995) 
and some have suggested that it also happens in AGNs (Storchi--Bergmann 1991; 
Storchi-Bergmann \& Wilson 1996; Ohyama et al. 1997). 

Another possibility is that SBNGs have an extra source of excitation. This, in 
fact, was the solution preferred by Kennicutt et al. (1989), who suggested a 
hidden AGN as the extra source. As proposed by Lehnert \& Heckman (1994), 
shocks produced by supernovae or starburst winds could probably also do the 
trick.  However, the extra excitation hypothesis is in contradiction with the 
use of emission--line diagnostic diagrams to classify activity types in 
galaxies. Indeed, we remind the reader that we already eliminated from our 
sample all the cases which appeared ambiguous in these diagnostic diagrams. 
Therefore, if the excess emission observed in these galaxies is still due to a residual 
excitation effect, it would seriously question the use of these diagnostic 
diagrams for distinguishing between the different activity types in 
emission--line galaxies. At the present time however, there is no evidence 
that diagnostic diagrams can give a wrong classification (BPT; VO; 
Gon\c{c}alvez et al. 1998; Kim et al. 1998). 

As an alternative, Shields \& Kennicutt (1995) have suggested that metal 
depletion occurring during dust formation could produce a slight extra 
excitation. At first, this seems perfectly applicable to the FIR-bright SBNGs. 
However, the fact that we do not observe any difference in the diagnostic 
diagrams between these dusty galaxies and their UV-bright counterparts 
suggests that dust does not play a significant role in producing the 
excess of emission in these galaxies. This interpretation is consistent with 
the original analysis made by VO on the null effect of dust on the different 
line ratios used in the diagnostic diagrams. It is also consistent with the 
fact that the two types of SBNGs have similar properties in the far-infrared 
(Coziol et al. 1998a). 

If the excess emission in the SBNGs with respect to normal disk H\,{\sc ii} 
regions is due to a residual excitation effect, we cannot use the empirical 
method devised by TEH to determine the abundance of nitrogen in SBNGs. In 
fact, we would even caution against using this method 
to determine the nitrogen abundance, as TEH have done, in early-type spirals and the 
nuclei of normal galaxies. It is therefore important to test thoroughly the 
different hypotheses for the origin of the excess emission.     

%______________________________________________________________

\begin{table}
   \caption[]{Mean values of emission per level of excitation bins in SBNGs}
      \label{Table1}
   \[
\begin{array}{ccccc}
\hline
\noalign{\smallskip}
{\rm Center\ bin} & {\rm N} & \log({\rm [NII]/H}\alpha) & \log({\rm
[OIII]/H}\beta) &  \Delta({\rm [NII]/H}\alpha)\\
\noalign{\smallskip}
\hline
-1.0 &  22 &  -0.37\pm0.12 & -0.95\pm0.09 & +0.19 \\
-0.6 &  87 &  -0.39\pm0.11 & -0.58\pm0.10 & +0.09 \\
-0.2 &  72 &  -0.42\pm0.11 & -0.21\pm0.12 & +0.16 \\
+0.2 &  52 &  -0.68\pm0.20 & +0.19\pm0.11 & +0.20 \\
+0.6 &  10 &  -1.05\pm0.16 & +0.53\pm0.08 & +0.21 \\
\noalign{\smallskip}
\hline
\noalign{\smallskip}
{\rm Center\ bin}& {\rm N}    & \log({\rm [SII]/H}\alpha) & \log({\rm
[OIII]/H}\beta) &  \Delta({\rm [SII]/H}\alpha)\\
\noalign{\smallskip}
\hline
-1.0  &  22  &  -0.63\pm0.11  &  -0.94\pm0.08  & +0.15 \\
-0.6  &  87  &  -0.60\pm0.10  &  -0.58\pm0.11  & +0.07 \\
-0.2  &  72  &  -0.57\pm0.09  &  -0.21\pm0.12  & +0.13 \\
+0.2  &  52  &  -0.58\pm0.14  &  +0.19\pm0.11  & +0.21 \\
+0.6  &  10  &  -0.72\pm0.14  &  +0.53\pm0.12  & +0.25 \\
\noalign{\smallskip}
\hline
\noalign{\smallskip}
{\rm Center\ bin}& {\rm N}    & \log({\rm [OI]/H}\alpha) & \log({\rm
[OIII]/H}\beta) & ---\\
\noalign{\smallskip}
\hline
-1.0  &   9  &  -1.60\pm0.18  &  -0.91\pm0.06 & ...\\
-0.6  &  49  &  -1.60\pm0.27  &  -0.57\pm0.10 & ... \\
-0.2  &  49  &  -1.45\pm0.18  &  -0.22\pm0.12 & ...\\
+0.2  &  21  &  -1.56\pm0.16  &  +0.18\pm0.11 & ... \\
+0.6  &   5  &  -1.61\pm0.14  &  +0.52\pm0.08 & ...\\
\noalign{\smallskip}
\hline
\end{array}
\]
\end{table}
%______________________________________________________________

\subsection{Excitation effects other than O and B stars} 

If the excess emission observed in SBNGs is caused by a surplus of excitation, 
the intensities of the emission lines should be correlated; they should all 
increase simultaneously.  To look for such a correlation, we have computed the 
mean values of [N\,{\sc ii}], [S\,{\sc ii}] and [O\,{\sc i}]  (relative to 
H$\alpha$) in excitation bins of 0.4 dex. These mean values are given in 
Table~1 and plotted in Fig~\ref{Fig.1}c, f and i. Because we did not find any 
difference between the spectral characteristics of the FIR- and UV-bright 
SBNGs, we merged the two samples together for this analysis. 

The mean excess of emission with respect to normal disk H\,{\sc ii} regions is 
about the same for nitrogen and sulphur, $+0.17\pm0.06$ dex and $+0.16\pm0.08$ 
dex respectively. However, the difference in [N\,{\sc ii}]/H$\alpha$ stays 
almost constant while the difference in [S\,{\sc ii}]/H$\alpha$ increases with 
the level of excitation (last column of Table~1). 

It also appears that [N\,{\sc ii}]/H$\alpha$ increases as the level of 
excitation decreases (or as the oxygen abundance increases), while [S\,{\sc 
ii}]/H$\alpha$ and [O\,{\sc i}]/H$\alpha$ stay constant. In Fig~\ref{Fig.1}c, f 
and i, it can be seen that [N\,{\sc ii}]/H$\alpha$ is correlated with [O\,{\sc 
iii}]/H$\beta$, while [S\,{\sc ii}]/H$\alpha$ and [O\,{\sc i}]/H$\alpha$ are 
not. This is confirmed with a confidence limit higher than 99\% using a Generalized 
Kendall's tau correlation coefficient test. 

These different behaviors of the emission-line ratios with the level of 
excitation are inconsistent with the hypothesis that the excess of emission in 
SBNGs is produced by an additional excitation effect.  On the other hand, they 
are partly consistent with what is expected if these line ratios are sensitive 
to a variation of abundance. If nitrogen is produced by intermediate-mass 
stars while sulphur is produced by massive stars, the intensities of these two 
lines should not be correlated. The different behaviors of these two lines 
also suggest that, in SBNGs, the abundance of nitrogen will increase with that 
of oxygen while the abundance of sulphur will stay constant, as is observed 
in normal disk H II regions (Torres--Peimbert et al.\ 1989). We will 
show later (Sect.5.1) that these behaviors are consistent with a scenario of 
successive bursts for the chemical evolution of SBNGs. 

\begin{figure*}
\hskip -2cm 
\resizebox{16cm}{!}{\includegraphics{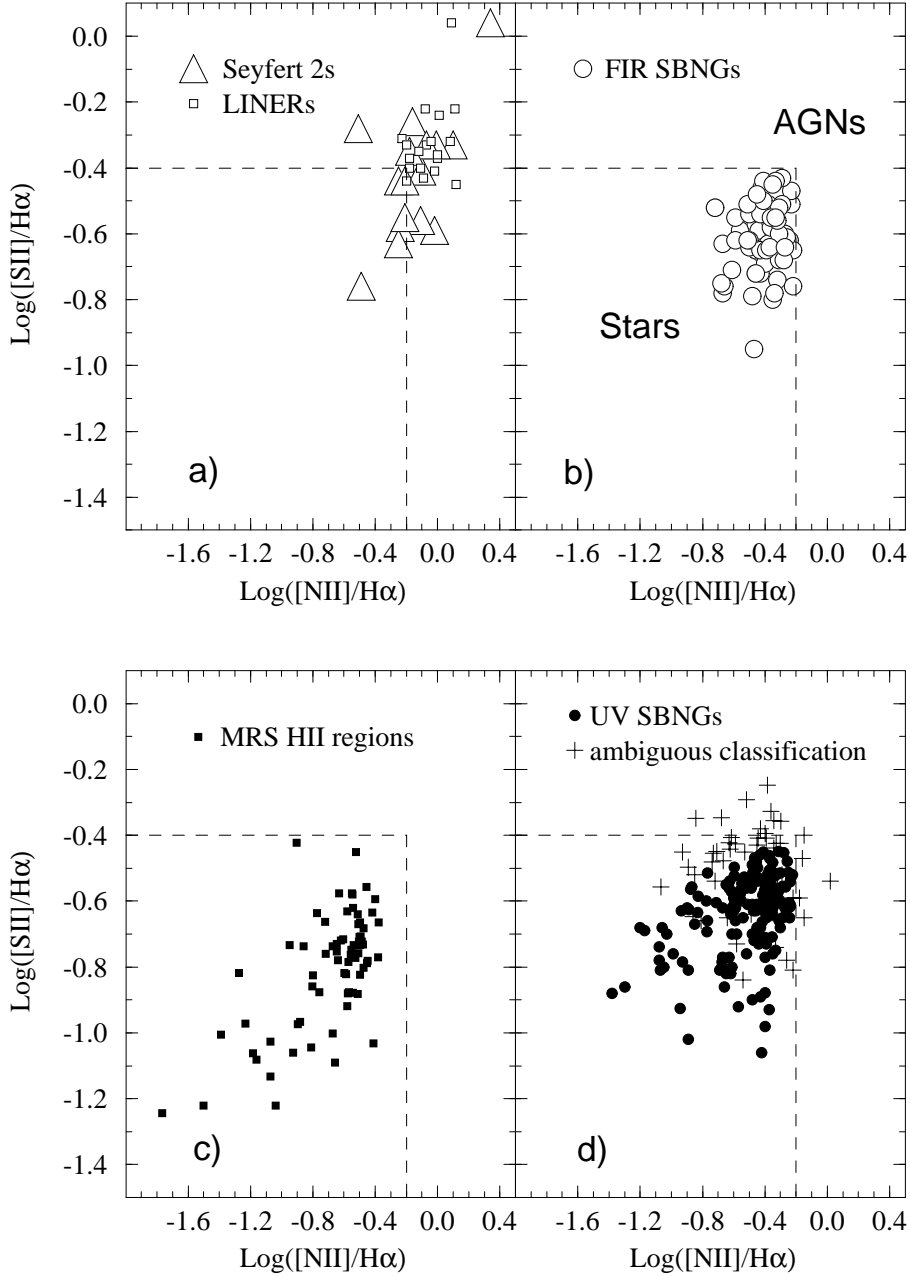}}
\hfill
\centering
\vskip -4cm
\parbox[b]{18cm}{
\caption[h]{Diagram of [S\,{\sc ii}]/H$\alpha$ as a function of [N\,{\sc 
ii}]/H$\alpha$ for a) the FIR-bright Seyfert 2s and LINERs; b) the FIR-bright 
SBNGs; c) the MRS sample of disk H\,{\sc ii} regions; d) the UV-bright SBNGs. 
The vertical and horizontal dashed lines separate the regions where AGNs
dominate over O and B stars as the main source of ionization of the 
gas} 
\label{Fig.2}} 
\end{figure*}

To better understand the difference between SBNGs and AGNs, we compare  
[N\,{\sc ii}]/H$\alpha$ with [S\,{\sc ii}]/H$\alpha$ in our different samples.
The two line ratios are significantly higher in the LINERs and Seyfert 2s than 
in the SBNGs (see Fig.~\ref{Fig.2}a and b). Note that we cannot distinguish 
between the two AGN types in this diagram. The SBNGs, on the other hand, are 
clearly separated from LINERs. Some Seyfert 2s have [S\,{\sc ii}]/H$\alpha$ 
ratios comparable to those of SBNGs, but this is not a problem because Seyfert 
2s can be clearly distinguished from SBNGs based on their higher excitation 
level in Fig.~\ref{Fig.1}. 

%----------------------------------------------------------- S_vib
%
%-----------------------------------------------------------

Using the positions of the SBNGs and Seyfert 2s/LINERs in Fig.~\ref{Fig.2}, we 
establish two regions in this diagram where the gas is excited by different 
mechanisms: AGN vs photoionization by O and B stars. This separation 
is also consistent with the lower limits proposed by Lehnert \& Heckman (1996) 
for the presence of diffuse ionized gas in the halos of edge-on starbursts.  
We test our separation criterion on the UV-bright SBNGs and on the MRS sample 
of disk H\,{\sc ii} regions in Fig.~\ref{Fig.2}c and d. As expected, all the 
H\,{\sc ii} regions in the MRS sample and the UV-bright SBNGs are located in 
the region where excitation by O and B stars is predominant. 

We also show in Fig.~\ref{Fig.2}d the position of the H\,{\sc ii} 
regions in our sample of SBNGs which were rejected on the basis of an 
ambiguous classification  in the three diagnostic diagrams.  Most of these 
ambiguous cases are either in the AGN regions or at the 
border of these regions. This supports our conclusion that the main 
source of excitation of the gas in SBNGs is O and B stars.   

\subsection{Possible spurious effects} 

\begin{figure*}
\hskip -2cm 
\resizebox{16cm}{!}{\includegraphics{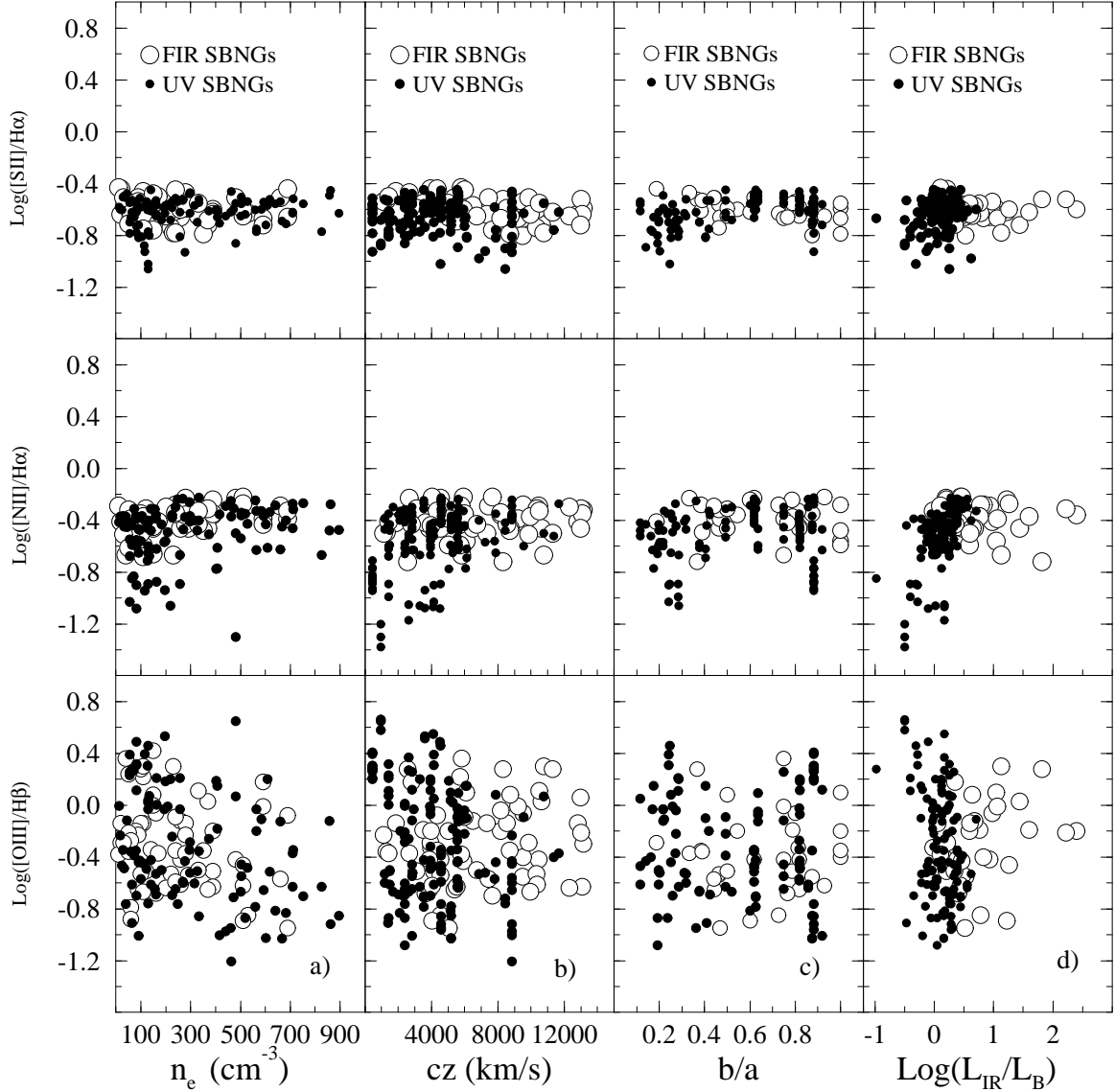}}
\hfill
\centering
\vskip -6.1cm
\parbox[b]{18cm}{
\caption{The variation of the three line ratios measured in the SBNGs as a 
function of a) the electronic density of the gas; b) the size of the 
regions covered by the slit, as represented by the redshift; c) the 
inclination of the galaxies given by the axis ratio b/a; d) the intensity 
of the burst, as indicated by the ratio L$_{IR}$/L$_B$. No difference is 
observed between the UV- and FIR-bright SBNGs. In general, the three line 
ratios are independent of these parameters} 
\label{Fig.3}} 
\end{figure*}

For our analysis to be complete, we now verify that no other factors influence 
the line ratios under investigation. The sulphur lines, for example, are 
sensitive indicators of electronic density, and one could imagine that a 
variation of this density over the line of sight introduces spurious effects. 
However, as shown in Fig.~\ref{Fig.3}a, the three line ratios are independent 
of electronic density. 

%----------------------------------------------------------- S_vib
%%______________________________________________________________ 

Another spurious effect will arise if the slit of the spectrograph covers 
regions with different physical properties, e.g. a central AGN and a 
circumnuclear starburst region, producing intermediate values for the line 
ratios.  The presence of such an effect can be tested by its dependence on 
redshift. If the physical conditions (different sources of excitation or 
density layers) change with radial distance, while the slit aperture remains 
constant, the line ratios will change with redshift. For example, an AGN in 
the nucleus of each SBNG would produce higher line ratios at low redshifts, 
because of the increasing contribution of the AGN over that of circumnuclear 
H\,{\sc ii} regions (Storchi-Bergmann 1991). The contrary is expected for 
diffuse gas (Lehnert \& Heckman 1994, 1996); the ratios would then increase 
with redshift due to the enhanced contribution of diffuse gas in the aperture. 
None of these effects is observed in Fig.~\ref{Fig.3}b. 

We also checked if the line ratios vary with the inclination of the galaxies 
(as estimated by their axis ratio {\it b/a}); this is shown in 
Fig.~\ref{Fig.3}c. Lehnert \& Heckman (1996) found different variations of the 
line ratios along the two axes, which they attributed to the presence of 
diffuse ionized gas in the halos of some FIR-bright starbursts. Again, no 
significant effect is present in the two samples of SBNGs. 

\begin{table}
\caption[]{Model for N159 in the LMC}
\[
\begin{array}{lcccc}
\hline
\noalign{\smallskip}
{\rm Ion} & \lambda & {\rm Obser.} & {\rm Model} & {\rm Model^{\mathrm{a}}}\\
\noalign{\smallskip}\hline\noalign{\smallskip}
{\rm CIII}     & 1909 & 0.103 & 0.103 & 0.518 \\
{\rm [OII]}   & 3727 & 1.235 & 1.236 & 1.752 \\
{\rm [NeIII]}  & 3869 & 0.264 & 0.259 & 0.350 \\
{\rm H}\gamma      & 4340 & 0.412 & 0.404 & 0.476 \\
{\rm [OIII]}   & 4363 & 0.028 & 0.031 & 0.036 \\
{\rm H}\beta       & 4861 & 1.000 & 1.000 & 1.000 \\
{\rm [OIII]}   & 4959 & 1.427 & 1.483 & 1.456 \\
{\rm HeI}    & 5876 & 0.142 & 0.142 & 0.114 \\
{\rm H}\alpha      & 6563 & 3.997 & 3.980 & 2.833 \\
{\rm [NII]}   & 6584 & 0.206 & 0.206 & 0.146 \\
{\rm [SII]}   & 6716 & 0.175 & 0.173 & 0.120 \\
{\rm [SII]}   & 6731 & 0.132 & 0.131 & 0.091 \\
{\rm [ArIII]}  & 7135 & 0.177 & 0.176 & 0.116 \\
{\rm [OII]}   & 7325 & 0.101 & 0.061 & 0.039 \\
\noalign{\smallskip}\hline 
\end{array}
\]
\begin{list}{}{}
\item[$^{\mathrm{a}}$] Unreddened line intensities using the curve for LMC of Pei (1992)
\end{list}
\end{table}

Finally, we looked for a relation between the emission-line ratios and the 
intensity of the burst, as measured by the ratio L$_{IR}$/L$_B$.  Following 
Lehnert \& Heckman (1996), the most straightforward prediction of the 
superwind theory is a relation between the emission-line ratios and the star 
formation rates. No relation exists between the three line ratios and the 
intensities of the bursts in the FIR-bright SBNGs; this is shown in 
Fig.~\ref{Fig.3}d. The [N\,{\sc ii}]/H$\alpha$ ratio increases slightly with  
L$_{IR}$/L$_B$ in the UV-bright SBNGs, but this is because [N\,{\sc 
ii}]/H$\alpha$ is lower in the SBNGs with high excitation, which also happen to 
have the lowest star formation rates.     
 
All these negative results indicate that no spurious effects invalidate the 
results of our analysis of the emission-line ratios in SBNGs.  We thus 
conclude that there is no extra excitation effect in these galaxies.
We are therefore left with only one 
possible explanation for the origin of the excess of nitrogen emission in SBNGs, namely 
that they have different abundances than H\,{\sc ii} regions in normal  
spiral galaxies (Stauffer 1982). This is what we will now verify. 

\subsection{The excess of nitrogen emission as an abundance effect} 

\begin{figure}
\hskip -3.8cm 
\resizebox{14cm}{!}{\includegraphics{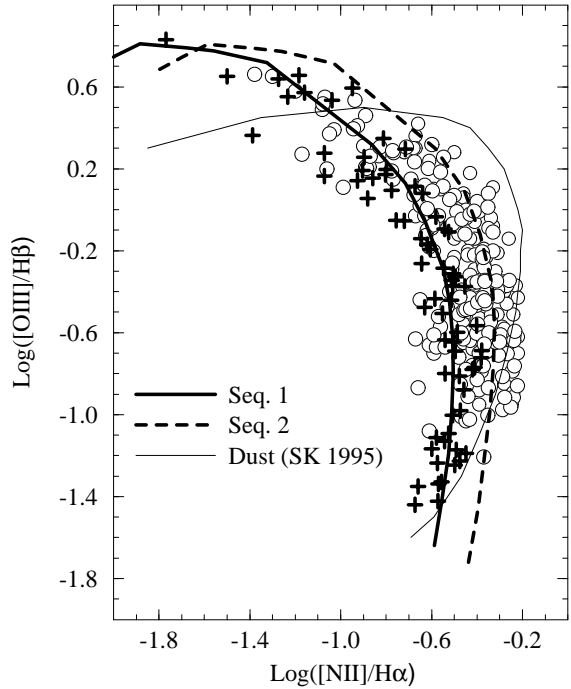}}
\hfill
%\centering
\vskip -9.5cm
\parbox[b]{8.8cm}{
\caption{Results of our photoionization sequences based on CLOUDY. Sequence~1 (Seq.~1)
reproduces the sequence traced by normal H\,{\sc ii} regions (thick crosses). 
Sequence~2 (Seq.~2) has twice as much nitrogen in abundance and reproduces the sequence traced by
the FIR- and UV-bright SBNGs (open circles). 
The thin continuous curve (identified as SK 1995) is the 
dust depletion model of Shields \& Kennicutt (1995).} 
\label{Fig.4}} 
\end{figure}

In their analysis of photoionized models of H\,{\sc ii} regions, Evans \&
Dopita (1985) pointed out that the position occupied by a galaxy in the 
diagnostic diagram [O\,{\sc iii}]/H$\beta$ vs [N\,{\sc ii}]/H$\alpha$ 
should also depend on the ratio N/O. It seems therefore reasonable to think that the 
higher [N\,{\sc ii}]/H$\alpha$ ratios in the SBNGs are due to an abundance 
effect. Indeed, a higher dispersion in N/O for a given 
metallicity will automatically translate into a higher dispersion 
of [N\,{\sc ii}]/H$\alpha$ for a given 
level of excitation, that is for a fixed [O\,{\sc iii}]/H$\beta$. 
To test this hypothesis we constructed two different H\,{\sc ii} 
region sequences using the photoionization code CLOUDY (Ferland 1997). 
For temperature $\le 50 000$\ K, we used Kurucz atmospheres, while for
higher temperature, the non--LTE model of Rauch (1997) was adopted. 
The resulting sequences are shown in Fig.\ref{Fig.4}.

Our first sequence (Seq.~1) was built to reproduce the spectral sequence for 
normal H\,{\sc ii} regions represented by the MRS sample. To constrain the 
sequence at the low-metallicity end, we used the parameters 
(Table~2) and abundances (Table~3) of the LMC H\,{\sc ii} region N159 
(observed by Dufour et al. 1982), deduced from a previous model also based on 
CLOUDY (Carlos Reyes et al. 1998). This reference point is crucial because the 
models of MRS and Dopita \& Evans (1986) do not reproduce the observations in 
this part of the diagnostic diagram. At a low metallicity level,  
we assumed that H\,{\sc ii} regions are photoionized by very young and hot stars.

\begin{table}
\caption[]{Abundances and Physical Parameters for N 159}
\[
\begin{array}{lcl}
\hline
\noalign{\smallskip}
        & {\rm Abundances}   & {\rm Physical Parameters}\\
{\rm X} & 12 +\log({\rm X/H}) &                         \\
\noalign{\smallskip}\hline\noalign{\smallskip}
{\rm He} &10.93  & \log({\rm L/L}_\odot)\dotfill 6.90 \\
{\rm C } & 7.86  & \log({\rm T}_{\rm eff}) ({\rm K})\dotfill 4.70 \\
{\rm N}  & 7.04  & \log({\rm n}_{\rm H}) ({\rm cm}^{-3})\dotfill 1.86 \\
{\rm O}  & 8.17  & \log(\tau)\dotfill 0.57 \\
{\rm Ne} & 7.46  & {\rm filling\ factor}\dotfill 0.03 \\
{\rm S}  & 6.66  & \log({\rm r}_{\rm in}) ({\rm cm})\dotfill 16.5  \\
{\rm Ar} & 5.94  & \log({\rm r}_{\rm ext}) ({\rm cm})\dotfill 20.0  \\
\noalign{\smallskip}\hline 
\end{array}
      \]
\end{table}
\begin{table}
\caption[]{Abundance ratios}
\[
\begin{array}{lccc}
\hline
\noalign{\smallskip}
{\rm Object}  & \log({\rm C/O}) & \log({\rm N/O}) & \log({\rm O/H}) + 12  \\
\noalign{\smallskip}\hline\noalign{\smallskip}
{\rm SMC^{\mathrm{a}}} & -0.8 & -1.4 & 7.96  \\
{\rm LMC^{\mathrm{a}}} & -0.4 & -1.0 & 8.40  \\
{\rm Sun^{\mathrm{b}}} & -0.3 & -0.9 & 8.84  \\
\noalign{\smallskip}\hline 
\end{array}
      \]
\begin{list}{}{}
\item[$^{\mathrm{a}}$] {\rm Carlos\ Reyes\ et\ al.\ (1998)}
\item[$^{\mathrm{b}}$] {\rm Grevesse\ \&\ Noels\ (1993)}
\end{list}
\end{table}

Starting from this point, we reproduced the spectral sequence for normal 
H\,{\sc ii} regions by varying the physical parameters and chemical 
abundances. For each abundance value, we searched for the combination of 
luminosity and ionization temperature that best reproduces the spectral 
sequence of H\,{\sc ii} regions. Except for He, C and N, the abundances 
changed linearly. The abundance of He was kept constant, while the ratios C/O 
and N/O varied following Eqs. (1) and (2). 

{\begin{eqnarray} 
\indent \log({\rm C/O}) = & -40.46 + 8.87[\log({\rm O/H}) + 12] \nonumber \\
            & - 0.49[\log({\rm O/H}) +12]^2  
\end{eqnarray} }

\begin{eqnarray} 
\indent  \log({\rm N/O}) = & 12.25 - 3.77[\log({\rm O/H}) + 12] \nonumber \\
            &+ 0.26[\log({\rm O/H}) + 12]^2   
\end{eqnarray} 

These two parametric relations were obtained from a regression on data from 
the LMC, SMC and solar values (Table~4). Our Eq. (2) yields a N/O ratio for 
the LMC which is consistent with the value given by the $primary + 
secondary$ relation proposed by Vila--Costas \& Edmunds (1993; hereafter 
VC\&E).  In the diagnostic diagram, the position of the galaxies with solar 
abundance was determined using the metallicity calibration curve of Coziol et 
al. (1994).

%----------------------------------------------------------- S_vib
%
%______________________________________________________________

Sequence~2 was derived from Sequence~1 by increasing by a 
factor 2 the abundance of nitrogen predicted by Eq. 
(2). As can be seen in Fig.\ref{Fig.4}, 
Sequence~2 reproduces the distribution in SBNGs relatively 
well. Note that the metallicities are comparable to those found in Sequence~1. 
Therefore, for a given metallicity, this 
%%model 
sequence predicts an overabundance of nitrogen 
in SBNGs with respect to normal disk H\,{\sc ii} regions. 
For comparison, we have also drawn in Fig.\ref{Fig.4} the predictions 
made by the model of Shields \& Kennicutt (1995) which takes into 
account the depletion of metals in dust. This model 
generally overpredicts the excess of nitrogen emission in SBNGs.
Considering the relatively good fit of Sequence~2, we conclude that 
the excess of nitrogen emission in SBNGs can easily be reproduced by 
a slight overabundance of this element. 

\section{Empirical estimates of the oxygen and nitrogen abundances} 

In the previous section, we performed several tests to verify the legitimacy 
of applying to the SBNGs the method for determining the nebular abundance in 
normal H\,{\sc ii} regions. First, we identified the main source of excitation 
of the gas in SBNGs with O and B stars. Then by constructing a simple 
%% model 
sequence of 
H\,{\sc ii} regions, we showed that the simplest hypothesis for explaining the 
excess of nitrogen emission observed in these galaxies is a slight 
overabundance of this element with respect to normal disk H\,{\sc ii} regions. 
We now feel justified to determine the oxygen and nitrogen abundances in our 
sample of SBNGs with the empirical techniques developed for normal H\,{\sc ii} 
regions. In doing so, we should also verify the predictions of our 
photoionized model, by comparing the abundances in the SBNGs with those of 
normal disk H\,{\sc ii} regions. 

\subsection{Abundances determination}

The oxygen abundances 
are determined following the relation (Vacca \& Conti 1992): 

\begin{eqnarray}
\indent \log({\rm O/H}) = -0.69\times \log({\rm R}_3) - 3.24
\end{eqnarray}

\noindent where R$_3 = 1.35\times$([O\,{\sc iii}]$\lambda5007/{\rm H}\beta)$. 

To determine the N/O ratios, we use the TEH method, adopting their empirical 
equation to estimate the electron temperature $t_2$ in the [N\,{\sc ii}] 
emission region. Using simple algorithms based on the three-level-atom solutions of McCall 
(1984), the abundance ratio N/O, assumed to be equal to N$^+$/O$^+$, is 
determined following the relation: 

\begin{eqnarray} 
\indent \log({\rm N/O})  = & \log \frac{6548+6584}{3726+3729} + 0.307-\frac{0.726}{t_2} \nonumber \\
                            &  -0.02\log(t_2) 
\end{eqnarray} 

\noindent where $t_2$ is in units of $10^{4}$\ K.

%% \begin{eqnarray} 
%% \log (N/O) = log \frac{6548+6584}{3726+3729} + 0.307-\frac{0.726}{t_2} 
%%                             -0.02\log(t_2) 
%% \end{eqnarray} 

For some of the galaxies in our sample, [O\,{\sc ii}]$\lambda3727$ was 
estimated using the empirical relation between [O\,{\sc 
ii}]$\lambda3727/$H$\beta$ and [O\,{\sc iii}]$\lambda(4959+5007)/$H$\beta$ 
proposed by MRS. Using their sample of galaxies, we verified that the N/O 
ratios obtained using the observed and estimated values of [O\,{\sc ii}] are 
the same. We also tested this relation for the SBNGs using the galaxies in our 
sample where [O\,{\sc ii}]$\lambda3727$ was directly measured. 

\subsection{The N/O ratio in normal H II regions} 

%----------------------------------------------------------- S_vib
\begin{figure}
%\vskip -2cm
\hskip -4.5cm 
\resizebox{14cm}{!}{\includegraphics{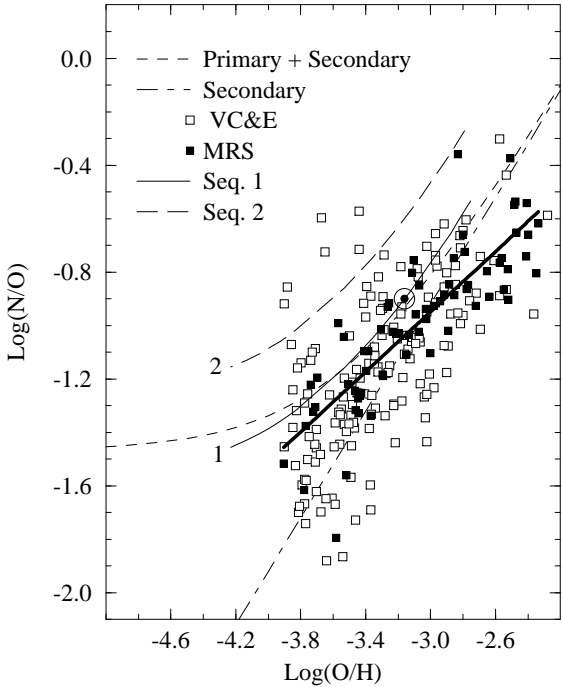}}
\hfill
%\centering
\vskip -9.5cm
\parbox[b]{8.8cm}{
\caption{Nitrogen vs oxygen abundances in normal H\,{\sc ii} regions.  
The predictions of our two sequences are compared with the 
N/O estimates in the MRS sample and the estimates for 
a sample of H\,{\sc ii} regions from the compilation of VC\&E.
The thick straight line is a linear fit to the MRS 
data. This line is offset by $\sim\ - 0.15$ dex relative to Sequence~1, 
suggesting that normal H\,{\sc ii} regions generally have subsolar N/O ratios} 
\label{Fig.5}} 
\end{figure}
%
% Regression lineaire sur MRS:
% log(N/O) = 0.56 +/- 0.04 log(O/H) + 0.74 +/- 0.14
% coefficient de correlation: 0.84
% Moyennes: O/H = -3.07 ; N/O = -0.99
%______________________________________________________________

We compare in Fig.~\ref{Fig.5} the predictions of Sequences~1 and 2 with the N/O 
estimates for the MRS sample of normal H\,{\sc ii} regions as well as for a 
subset of H\,{\sc ii} regions from the compilation of VC\&E.  The N/O 
estimates for the MRS sample of galaxies are $\sim\ 0.15$ dex lower, on 
average, than the values predicted by Sequence~1. In Fig.~\ref{Fig.5}, we also 
plotted the fits of the different models ($secondary$ and $primary + 
secondary$ production of nitrogen) proposed by VC\&E to explain the values 
observed in different normal H\,{\sc ii} regions. In general, the model for $secondary$ 
production seems to better describe the behavior of normal H\,{\sc ii} 
regions, but it underestimates the N/O ratio at solar oxygen abundance. 

%----------------------------------------------------------- S_vib
\begin{figure}
%\vskip -2cm
\hskip -4.5cm 
\resizebox{14cm}{!}{\includegraphics{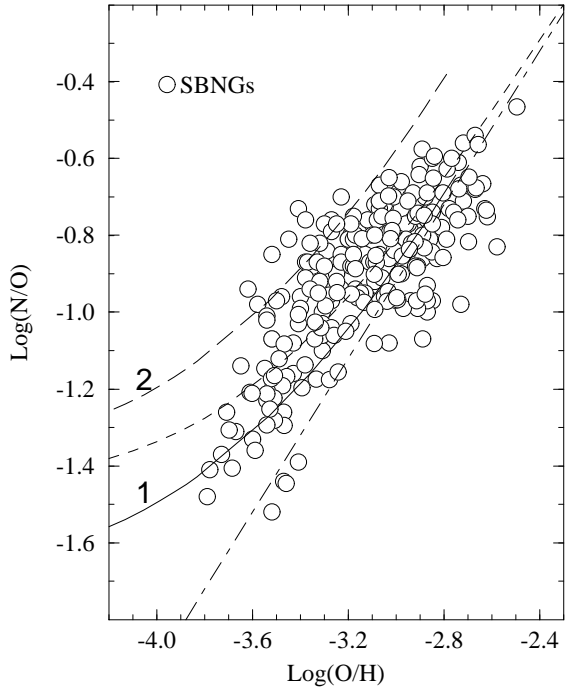}}
\hfill
%\centering
\vskip -9.5cm
\parbox[b]{8.8cm}{
\caption{Nitrogen vs oxygen abundances for the SBNGs. The $secondary$ 
(short-long dashed line) and $primary + secondary$ (dashed line) relations 
are also shown together with Sequences~1 (continuous line) and 2 (long dashed 
line) displaced by -0.15 dex 
in N/O to fit the position of normal disk H\,{\sc ii} regions.  
The SBNGs do not follow
the secondary relation. There is a sharp jump of $\log($N/O$) \sim 0.3$ at about 
$\log($O/H$) \sim - 3.4$ and the values almost level off 
in the range $-3.3 < \log($O/H$) <- 2.9$, which suggests 
that the increase of nitrogen abundance with that of oxygen is not a 
continuous process} 
\label{Fig.6}} 
\end{figure} 
% 
%______________________________________________________________ 

We verified that the offset of 0.15 dex between the data and Sequence~1 does not 
depend on the relation adopted for estimating the oxygen abundance. If we use 
the empirical equation proposed by Zaritsky et al. (1994) for example (or a 
similar relation deduced from our model) the oxygen abundances of the most 
metal-rich H\,{\sc ii} regions are slightly reduced, changing the slope of the 
MRS distribution and rendering it more similar to the $secondary$ relation. We 
could also add a term in the equation giving the N/O ratio. Indeed, the 
complete equation (see Pagel et al. 1992, their eq. 9) contains another term 
which depends on density and temperature. But this additional term is two 
orders too low to explain the offset, even if we assume extreme physical 
conditions (temperatures lower than $10^3$\ K and densities well over 1000 
cm$^{-3}$). 
%----------------------------------------------------------- S_vib
\begin{figure*}
\vskip -0.5cm
\hskip 2cm 
\resizebox{14cm}{!}{\includegraphics{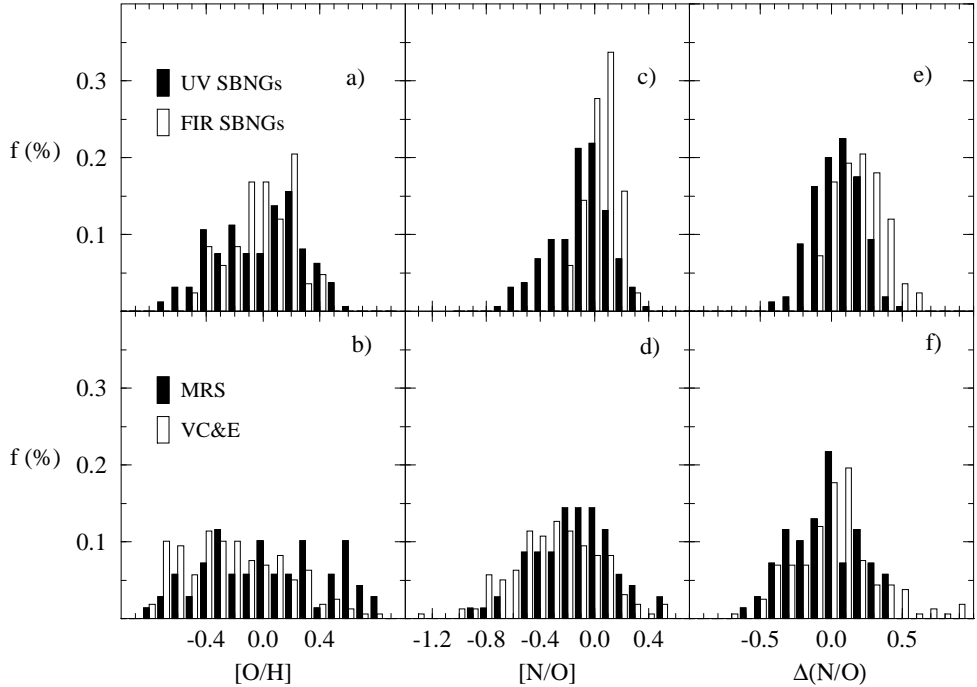}}
\hfill
%\centering
\vskip -8cm
\parbox[b]{18cm}{
\caption{Distribution of oxygen and nitrogen abundances, 
both relative to the solar value, and of the deviations $\Delta$(N/O) of the 
observed values from the $secondary$ relation for the different samples 
considered in this paper: a), c) and e), the SBNGs; b), d) and f) the samples 
of normal H\,{\sc ii} regions of MRS and VC\&E.  The SBNGs 
have oxygen abundances comparable to those of the disk H\,{\sc ii} regions, 
but a slight overabundance of nitrogen and do 
not follow the $secondary$ relation.} 
\label{Fig.7}} 
\end{figure*}
%
%______________________________________________________________

Although 0.15 dex is probably smaller than the uncertainties on our estimates, 
we believe that the offset of the MRS sample with respect to %% our model 
sequence 1 is 
real. In fact, MRS already observed such a difference with their own 
models\footnote{Instead of lowering the nitrogen abundance, they raised the 
oxygen abundance to 2.3 times solar to solve the problem. This is of the 
order of the offset observed in our model.}. 
In support of this hypothesis, we note that the 158 disk H\,{\sc ii} regions 
from the compilation of VC\&E in general follow the $secondary$ relation in 
Fig.~\ref{Fig.5}. The same result was found recently by van Zee et al. 
(1998). It seems, therefore, that normal H\,{\sc ii} regions usually have 
lower than solar N/O values at solar metallicity.  Consequently, the behavior 
of the H\,{\sc ii} regions in the MRS sample is normal.     

\subsection{The N/O ratio in SBNGs} 

Our estimates of the N/O ratios in SBNGs are shown in Fig.~\ref{Fig.6}.  In 
these galaxies, a clear excess of nitrogen with respect to normal H\,{\sc ii} 
regions is observed. For the sake of comparison, the predictions of our Sequences 
1 and 2 were plotted in Fig.~\ref{Fig.6} with an offset of $-0.15$ dex in N/O.
From this, we deduce that the abundance of nitrogen in the SBNGs is at most a 
factor two higher than in normal disk H\,{\sc ii} regions.  

Examining how the abundance of nitrogen varies with that of oxygen, we find 
that the SBNGs do not follow the $secondary$ relation.  This is shown in Fig.~\ref{Fig.6}. A linear fit  
yields (with a correlation coefficient of 76\%) $\log($N/O$) = 0.55\log($O/H$) + 0.8$, which  
is consistent with a mixture of $primary + secondary$ mode of production 
(McGaugh 1991). But the increase of the N/O ratio with metallicity does not
look like a continuous process. The N/O ratio rises 
sharply by about 0.3 dex at an oxygen abundance of $\sim -3.4$ and stays 
almost constant in the range $-3.4 <$log(O/H)$< - 2.9$.  It is important to 
remember that there are 243 points in this figure, therefore the lack of 
values on the $secondary$ relation over this particular range of metallicities 
cannot be attributed to an incompleteness of the data (compare also with 
Fig.~\ref{Fig.5} where we have a similar number of normal H\,{\sc 
ii} regions).  

In order to better understand what the excess of nitrogen abundance in SBNGs 
means, we compare the distributions of oxygen and 
nitrogen abundances (in percentages), both relative to their solar value. The 
distributions of oxygen abundances for the different samples, shown in
Fig.~\ref{Fig.7}, are similar. 
The samples cover almost uniformly a metallicity range between $-0.8$\ and 
$0.8$\ dex. The FIR-bright SBNGs seem slightly more metal-rich than all the 
other galaxies in our sample. 

The distributions of nitrogen abundances, on the other hand, are 
significantly different in SBNGs and in disk H\,{\sc ii} regions: a higher 
fraction of SBNGs have solar abundances. The dispersion is relatively 
high for normal disk H\,{\sc ii} regions and their distributions peak at a 
value $\sim -0.2$ dex. This is shown in Fig.~\ref{Fig.7}c and d. The 
dispersion of nitrogen abundances in the FIR-bright SBNGs is remarkably small.  
This constancy of N/O (around log(N/O) = -0.8) translates into the 
pseudo plateau seen in Fig.~\ref{Fig.6}.  

Next, we investigate the deviation $\Delta$(N/O) which is the difference 
between the observed nitrogen abundance and that predicted by the $secondary$ 
relation at the corresponding oxygen abundance.  As shown in Fig.~\ref{Fig.7}e 
and f, a high number of SBNGs have a positive deviation from the $secondary$ relation. This 
confirms the peculiar structures observed in our data in Fig.~\ref{Fig.6}.   

At first, it may seem surprising to find that what we define as ``normal'' 
H\,{\sc ii} regions have slightly lower N/O ratios than solar. But there is 
nothing which assures us that the solar value should be the standard in these 
objects. This result goes in the same sense as those based on studies of the 
metallicities of stars in the solar neighborhood: the most frequent 
metallicity (estimated here by [Fe/H]) is around $-0.15$ dex (Cayrel et al. 
1997) and the Sun seems to have a higher metallicity than the majority of 
stars (Eggen 1978).    

The fact that the N/O distribution for SBNGs peaks at the solar value implies 
a slight overabundance of nitrogen with respect to normal disk H\,{\sc ii} 
regions. But the range of N/O values found in the SBNGs is comparable to 
that observed in the bulges of normal early-type spiral galaxies (TEH; van Zee 
et al. 1998). This is consistent with the recent
discovery made by TEH. These authors found that H\,{\sc ii} regions in
early-type spirals have slightly higher N/O 
ratios than H\,{\sc ii} regions in late-type spirals.
Because the H\,{\sc ii} regions in late-type spirals are generally 
more numerous and luminous than in early-type spirals, samples of H\,{\sc ii} 
regions in normal galaxies are naturally biased towards those in late-type 
spirals, as can be easily verified for the MRS, VC\&E and van Zee et al. 
(1998) samples. The SBNGs, on the other hand, are more numerous among 
early-type spirals (see Coziol et al. 1998a), explaining the observed
higher abundance ratio in this sample.  

We conclude that our observations on the nitrogen abundance
of SBNGs are consistent with the chemical evolution of 
early-type spiral galaxies. It suggests that what we see could be the main 
production of nitrogen in the bulges of these galaxies. 
  
\section{Discussion} 

\subsection{The origin of nitrogen in SBNGs} 

It is usually assumed that nitrogen is the product of the evolution of 
intermediate-mass ($0.8$M$_\odot\leq $~M$\leq 8$M$_\odot$) stars. Consequently, 
the expected behavior of N/O as a function of O/H could be $primary$, 
$secondary$ or a mixture of both (van den Hoek \& Groenewegen 1997). But
a mixture of $primary + secondary$ mode of production could also imply a contribution by 
massive stars. These two possibilities have very different consequences for 
the nature of SBNGs. 

In the literature, we find some cases where a temporary contamination of 
nitrogen by massive stars is suspected (Pagel et al. 1992). The contamination 
is supposed to come from massive Wolf-Rayet stars. But evidence in favor of a 
significant contribution of nitrogen by massive stars is still weak (Esteban \& 
Peimbert 1995). In a recent paper, Kobulnicky \& Skillman (1998) showed that, 
even in H\,{\sc ii} galaxies, which can be considered younger than  
SBNGs, the nitrogen abundance seems to be correlated to the carbon abundance, 
which is inconsistent with a contribution by massive stars. 

For the more massive and evolved SBNGs, a significant contribution of nitrogen by massive 
stars would imply major bursts of star formation, not much older than a few 
Myrs. But, with the exception of the ultra-luminous infrared galaxies,  
observations of SBNGs do not support such a scenario. 
Their relatively low H$\alpha$ equivalent widths, reddish $(B-V)$ 
colors, L(H$\alpha$)/L(IR) and L(B)/L(IR) luminosity ratios are more 
consistent with low-level bursts occurring on a long or recurrent mode (Coziol 
1996).  In general, SBNGs also have relatively low 
rates of supernovae (Turatto et al. 1989; Richmond et al. 1998; 
Gonz\'alez~Delgado et al. 1999), 
which is inconsistent with very young and massive bursts. 

Another difficulty with the major burst scenario for SBNGs is 
to identify the origin of such sudden and energetic events. The only mean we 
know of is interaction (or merger) between two massive spiral galaxies, like  
the ultra-luminous infrared galaxies for example. But most SBNGs are 
relatively isolated and the origin of the bursts cannot be blamed on a recent 
interaction with a nearby massive companion (Coziol et al. 1997a, 1997b; 
Contini et al. 1998).      
       
The production of nitrogen by intermediate-mass stars is better documented and 
a more likely event.  According to van den Hoek \& Groenewegen (1997), the bulk of 
nitrogen in galaxies is produced by AGB stars during their thermal pulsing 
phase, when the stars lose most of their mass to the ISM ($\sim 0.4 $ 
M$_\odot$ for a 1 M$_\odot$ star and $\sim 4.8 $ M$_\odot$ for a 6 M$_\odot$ 
star). The low mass (M $\leq 4$ M$_\odot$) AGBs contribute to He and C, while 
the high mass AGBs contribute to He and N. The typical lifetime of a 4 
M$_\odot$ star is $\sim 2\times10^8$ yrs, while that of a 8 M$_\odot$ star is 
$\sim 4\times10^7$ yrs (Charlot \& Bruzual 1991). These time scales are much 
longer than the typical lifetimes of massive O and B stars.  Therefore,
following this scenario, the bursts originate much further back in 
time, which solves the problem of the relative isolation. 

According to the model of Charlot and Bruzual (1991), the early phase of a 
burst is dominated by massive stars of types O to B5. The AGB branch turns on after 
about 0.1 Gyr. Between 0.4 and 1.6 Gyrs, the AGB population accounts for $\sim 
35$\% of the bolometric luminosity, while massive stars of type B5 to A7 
contribute $\sim 40$\%. Between 1.6 and 8 Gyrs, the AGB contribution drops to 
$\sim 20$\%, and stars of type A7 to G0 dominate the main sequence. From this, 
we conclude that the main phase of production of nitrogen in a starburst 
should occur between 0.4 and 1.6 Gyrs after the beginning of the burst (this 
implies the evolution of stars from 8 to 3 M$_\odot$). Now Coziol (1996) found
that the star formation in SBNGs was nearly constant over the last 2-3 
Gyr period. Thus, if we assume a median age of 1 Gyr for the bursts, 2 or 3 bursts 
(or more if the bursts have shorter durations) could be sufficient to 
produce the number of co-evolved intermediate-mass stars needed for 
the abundance of nitrogen observed today. 

%----------------------------------------------------------- S_vib
\begin{figure}
\vskip -0.25cm
\hskip -4.2cm 
\resizebox{14cm}{!}{\includegraphics{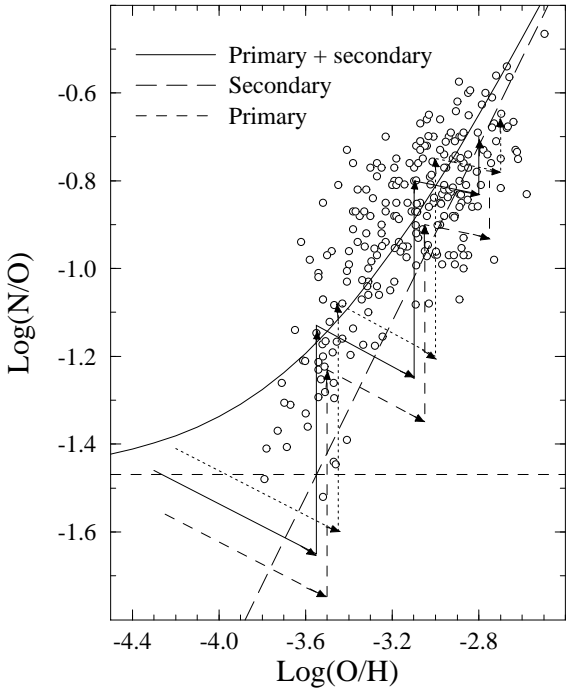}}
\hfill
%\centering
\vskip -9.5cm
\parbox[b]{8.8cm}{
\caption{Schematic representation of the process of production of nitrogen in 
SBNGs in a sequence of bursts, based on the analytical model 
presented by Garnett (1990). This model assumes that each burst produces an 
oxygen enrichment due to the evolution of massive stars followed by a 
significant nitrogen enrichment which comes from the evolution of intermediate-mass 
stars. If one assumes multiple bursts of decreasing intensity, the sum of the 
vectors converges to a line whose slope corresponds to the mean increase of 
chemical abundance in time (the $secondary$ relation). 
The dispersion is related to different initial 
intensities or different ages (represented by only three different vector sums 
on the figure). The open circles represent the two samples of SBNGs} 
\label{Fig.8}} 
\end{figure}
%
%______________________________________________________________

In Fig.~\ref{Fig.8}, we show schematically how a sequence of bursts could 
explain the production of nitrogen in the SBNGs. Our scenario is based on the 
analytical model presented in Garnett (1990). We assume that SBNGs begin their 
chemical evolution with N/O and O/H ratios typical of H\,{\sc ii} galaxies. 
The first phase in their chemical evolution corresponds to an increase of O/H 
due to the evolution of massive stars. At the same time N/O decreases (Garnett 
1990; Olofsson 1995). The second phase begins $\sim 0.4$ Gyrs after the onset 
of the burst with the evolution of intermediate--mass stars. Because these 
stars produce mostly nitrogen, only N/O increases during this phase.  

The second burst begins with an increase in O/H, but with a lower amplitude 
than in the first burst.  Indeed, models of sequential bursts in galaxies 
usually assume that the successive bursts will have decreasing intensities 
(Gerola et al. 1980; Kr\"{u}gel \& Tutukov 1993; Marconi et al. 1994: 
K\"{o}ppen et al. 1995). The decrease in N/O during oxygen enrichment (the 
slope of the vector) is also smaller as it becomes more difficult to lower 
this ratio as the oxygen abundance increases (Garnett 1990). Again, about 0.4 
Gyr after the beginning of the second burst, the intermediate stars evolve and 
N/O increases, although not as much as in the first burst. 

The norms of the vectors in Fig.~\ref{Fig.8} are related to the intensity of 
the bursts; the stronger the burst, the higher the increase in O/H and N/O. The 
difference between the norm of the oxygen and nitrogen enrichment vectors 
should depend on the IMF and on some other parameters related to the yields 
of the different stars. 

If we increase the number of bursts and assume that successive bursts
are weaker and weaker, the sum of the vectors converges 
towards a line whose slope represents the mean increase of O/H and N/O 
in time, or in other words towards the $secondary$ relation. Following this 
scenario, a constant star formation can be viewed as an infinite sum of very 
low-intensity bursts of star formation, explaining why normal disk H\,{\sc ii} 
regions follow the $secondary$ relation. 

%----------------------------------------------------------- S_vib
\begin{figure}
\vskip -0.25cm
\hskip -4.2cm 
\resizebox{14cm}{!}{\includegraphics{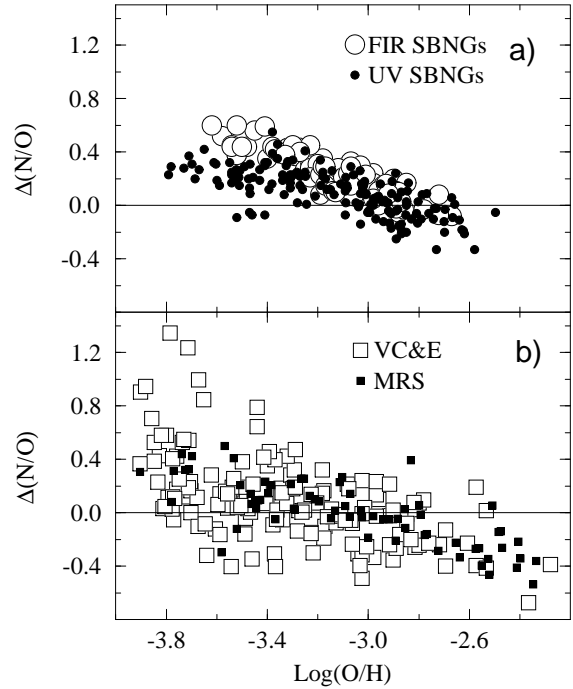}}
\hfill
%\centering
\vskip -9.5cm
\parbox[b]{8.8cm}{
\caption{Deviation of the observed nitrogen abundance 
from the $secondary$ relation ($\Delta$(N/O)) vs oxygen abundance :
a) in the SBNGS, b) in the normal H\,{\sc ii} regions.
One prediction of the multiple burst model is that $\Delta$(N/O) 
decreases with time, or, equivalently, with increasing O/H.
This effect is present in the SBNGs; continuous star formation, like in 
normal disk H\,{\sc ii} regions will produce a lower deviation scattered about 
the $secondary$ relation. The effect is also partly there in 
normal H{\sc ii} regions; their behavior becomes more starburst-like at 
low metallicities} 
\label{Fig.9}} 
\end{figure}
%
%______________________________________________________________

One prediction of this model is that the deviation of the observed N/O ratio 
from the $secondary$ relation ($\Delta$(N/O)) will decrease with the age of 
starburst galaxies, or, equivalently, with the increase in metallicity. In 
Fig.~\ref{Fig.9}a, we see that this prediction is satisfied in the SBNGs. We 
have verified that there are no relations between the deviation $\Delta$(N/O) 
and the ratios [N\,{\sc ii}]/H$\alpha$ and [S\,{\sc ii}]/H$\alpha$, which rules 
out a possible explanation of this behavior in terms of an excitation effect 
(indeed, following the excitation hypothesis the points with higher 
$\Delta$(N/O) would also have higher excess of emission in nitrogen and 
sulphur, which is not observed). 

The fact that the deviation in Fig.~\ref{Fig.9}a is mostly positive could be 
explained by the different durations of the chemical evolutionary phases. 
Because massive stars have very short lifetimes, the oxygen enrichment phase 
is almost instantaneous, and O/H rapidly reaches a maximum (the tip of 
each horizontal vectors in Fig.~\ref{Fig.8}). The lifetime of the stars 
producing nitrogen, on the other hand, spans a much larger range of values. 
The nitrogen enrichment probably increases rapidly at the beginning, but is 
prolongated over a longer period of time, as the lower-mass stars evolve. As a 
result, the top of each vertical vector in Fig.~\ref{Fig.8} will always be 
much more populated, this phase representing a natural stable mode in the 
starburst's evolution.  
  
The above scenario for the formation of nitrogen in SBNGs has some interesting
implications that can be further tested.  Assuming that, in SBNGs, 
carbon is mainly produced by intermediate--mass stars,
we should observe a deficiency of C with respect to N, because
the AGB stars that produce the carbon have longer
main sequence lifetimes than those producing the nitrogen.
Then, by comparing the abundance of C with that of N we could
determine the age of the sequence of
bursts, while by comparing the abundance of N to that of S,
we could determine the intensity and IMF of the bursts.
Of course, none of this will be observed if carbon is mainly
produced by massive stars (Maeder 1992), its abundance increasing
with metallicity, which is indeed 
observed in irregular galaxies (Garnett et al. 1995)
and in the disks of {\it normal} spiral galaxies (Garnett et al. 1999).

\subsection{The origin of starbursts as a cosmological event} 
 
We have shown in previous papers (Coziol et al. 1997b, 1998c), that the most 
plausible hypothesis for explaining the nearby SBNGs is that they are the 
result of mergers of small-mass and gas-rich systems, probably similar to the 
nearby H\,{\sc ii} galaxies, which took place sometime in the recent past. The 
nearly constant star formation rate over a few Gyr period (Coziol 1996) and 
now the time delay implied for the production of nitrogen by intermediate-mass 
stars both suggest merger ages of the order of 2 to 3 Gyrs. On a cosmological 
time scale, this corresponds roughly to redshifts z $\sim 0.2$ to 0.3 (using 
H$_o=75$ km s$^{-1}$ Mpc$^{-1}$ and assuming $\Omega = 0$) in the past. Our 
hypothesis for explaining the nearby SBNGs predicts a population of merging 
galaxies at intermediate redshifts. Therefore, it is probably not a 
coincidence that we do find a substantial increase in the number of irregular 
blue star forming galaxies (Broadhurst et al. 1992; Cowie et al. 1996; Fioc 
\& Rocca--Volmerange 1997; Lilly et al. 1998) at these redshifts. Some of 
these galaxies have characteristics which are consistent with our 
interpretation for the SBNGs, either in terms of star formation (Guzman et 
al. 1997; Glazebrook et al. 1998; Fioc \& Rocca--Volmerange 1997) or in 
terms of morphology (Colless et al. 1994; Abraham et al. 1996; van den Bergh 
et al. 1996; Teplitz et al. 1998). 

It is clear, on the other hand, that these pieces of evidence are not 
sufficient for identifying all the Faint Blue Galaxies as the merging 
progenitors of the SBNGs. There are still substantial problems to solve before 
understanding what the relation between these two phenomena could be. One of 
these problems, for instance, is the elapsed time between the two events. If 
some of the Faint Blue Galaxies merged to form the nearby SBNGs, they merged 
sometime between 2 or 3 Gyrs ago. These time scales are much too long compared 
to the predictions of the best models of interacting-merging galaxies that we 
can find in the literature (Mihos \& Hernquist 1994, 1996 for example). How 
could these galaxies sustain high rates of star formation over such long time 
scales? We already proposed one solution, which is that the 
merging phase produces more than one burst of star formation. This could 
probably be achieved by introducing in the interacting-merging models 
different initial conditions, which would better suit the characteristics 
of proto-galaxies (the merging galaxies
are not massive spirals but small-mass, irregular, metal-poor and gas-rich 
galaxies), and by regulating the bursts of star formation with a mechanism like 
the feedback of supernovae (Gerola et al. 1980; Kr\"{u}gel \& Tutukov 1993)
or by repeating them in a sequence of mergers (Tinsley \& Larson 1979).

\section{Summary and Conclusions} 

Using the same empirical methods as for normal H\,{\sc ii} regions, we have 
estimated the N/O and O/H ratios of a sample of 243 H\,{\sc ii} regions 
observed in different SBNGs. We have found clear evidence for an overabundance 
of nitrogen with respect to normal disk H\,{\sc ii} regions with comparable 
metallicities. The range of N/O values found in the SBNGs is comparable to 
that observed in the bulges of early-type spiral galaxies. Our observations 
are consistent with the discovery made by TEH that the N/O ratios of H\,{\sc 
ii} regions in the bulges of early-type spiral galaxies are higher than in 
late-type spirals. For the SBNGs, this suggests that what we see could be the 
main production of nitrogen in their bulges. 
    
The most plausible hypothesis for the origin of nitrogen in SBNGs is that it 
is the product of the evolution of intermediate-mass stars. These stars would 
have formed during a sequence of bursts extending over a few Gyr period. 
Following this scenario, the origin of the bursts is located 2 to 3 Gyrs in 
the past. On a cosmological time scale these time intervals correspond to 
redshifts z $\sim 0.2$ to 0.3, where one observes a substantial increase in the 
number of star forming galaxies whose characteristics are consistent with our hypothesis. 
This implies that some of the Faint Blue Galaxies observed at 
intermediate redshifts may have merged to produce the SBNGs observed today. 

The possible affiliation of the nearby SBNGs with the excess of blue galaxies 
in the field at intermediate redshifts suggests that the SBNGs are not a 
peculiar phase in the evolution of galaxies, but the result of a process which 
was much more common in the recent past of the Universe. This process we 
tentatively identify with hierarchical formation of galaxies. 
 
\acknowledgements 

We would like to thank Dr. C. Batalha who instructed us about the stellar 
abundance anomalies observed in the solar neighborhood and Dr. R. R. de 
Carvalho for very stimulating discussions on galaxies formation and cosmology. 
We thank the referee, Dr. M. Peimbert, for a critical reading and judicious 
suggestions.
We are also thankful to Dr. G. J. Ferland for kindly providing the code 
CLOUDY and to the technical staff of Observatoire de Haute-Provence for 
assistance at the telescope. Carlos Reyes acknowledges with thanks the receipt 
of research fellowships from the CNPq.

\end{document}